\documentclass[a4paper,11pt]{article}
\usepackage{pos}
\usepackage{graphicx}
\usepackage{dcolumn}
\usepackage{bm}
\usepackage{amsmath}
\usepackage{braket}
\usepackage{slashed}
\usepackage{epstopdf}
\usepackage{placeins}
\usepackage{multirow}
\usepackage{makecell}
\usepackage{soul}
\usepackage{braket}
\usepackage[shortlabels]{enumitem}
\usepackage{placeins}
\usepackage{adjustbox}
\usepackage[font=small, labelfont=bf, textfont={small,it}]{caption}
\usepackage{comment}
          
\title{Sphaleron rate as an inverse problem: a novel lattice approach}

\author[a]{Claudio Bonanno}
\author[b]{Francesco D'Angelo}
\author[b]{Massimo D'Elia}
\author[b]{Lorenzo Maio}
\author*[b,c]{Manuel Naviglio}

\affiliation[a]{Instituto de F\'isica Te\'orica UAM-CSIC, c/ Nicol\'as Cabrera 13-15, Universidad Aut\'onoma de Madrid, Cantoblanco, E-28049 Madrid, Spain}
\affiliation[b]{Dipartimento di Fisica dell'Universit\`a di Pisa \& \\ INFN Sezione di Pisa, Largo Pontecorvo 3, I-56127 Pisa, Italy}
\affiliation[c]{Scuola Normale Superiore, Piazza dei Cavalieri 7, I-56126 Pisa, Italy}

\emailAdd{manuel.naviglio@sns.it}

\abstract{We compute the sphaleron rate on the lattice. We adopt a novel strategy based on the extraction of the spectral density via a modified version of the Backus-Gilbert method from finite-lattice-spacing and finite-smoothing-radius Euclidean topological charge density correlators. The physical sphaleron rate is computed by performing controlled continuum limit and zero-smoothing extrapolations both in pure gauge and, for the first time, in full QCD.}

\FullConference{The 40th International Symposium on Lattice Field Theory (Lattice 2023)\\
July 31st - August 4th, 2023\\
Fermi National Accelerator Laboratory\\}

\begin{document}
\maketitle

\section{The Sphaleron Rate}
The strong sphaleron rate is defined as
\begin{equation}\label{eq:rate_def}
\begin{split}
\Gamma_{sphal}  = \underset{t_{\mathrm{M}}\to\infty}{\underset{V_s\to\infty}{\lim}} \, \frac{1}{V_s t_{\mathrm{M}}}\left\langle\left[\int_0^{t_{\mathrm{M}}} d t_{\mathrm{M}}' \int_{V_s} d^3x \, q(t_{\mathrm{M}}', \vec{x})\right]^2\right\rangle
=\int d t_{\mathrm{M}} d^3x \braket{q(t_{\mathrm{M}},\vec{x}) q(0,\vec{0})},
\end{split}
\end{equation}
where $t_\mathrm{M}$ is the Minkowski time and $q = (\alpha_s/8\pi)G\widetilde{G}$ is the QCD topological charge density. This quantity plays a very key role in different phenomenological situations.

Recently in~\cite{Notari:2022ffe} it has been suggested that the strong sphaleron rate plays a crucial role in axion phenomenology. Specifically, the QCD strong sphaleron rate governs the creation and annihilation of axions in the early Universe and directly influences the Boltzmann equation, which describes the time-evolution of the axion number distribution in the cosmological medium.
Furthermore, during heavy-ion collisions, a formation of a hot quark-gluon plasma with strong magnetic fields occurs. The non-zero sphaleron rate within this plasma can induce local imbalances in left and right-handed quark species, resulting in phenomena such as the Chiral Magnetic Effect. This effect manifests as an electric current parallel to the magnetic field within the quark-gluon medium, see~\cite{Fukushima:2008xe, Kharzeev:2013ffa, Astrakhantsev:2019zkr, Almirante:2023wmt} for details. 

This attention from different points of view calls for a fully non perturbative computation of the QCD sphaleron rate at finite temperature. Previously, other computations have been performed, although only restricted to pure gauge~\cite{Kotov:2018aaa, Kotov:2019bt, Altenkort:2020axj, BarrosoMancha:2022mbj}. In this paper we present our results in pure gauge for a temperature $T \simeq 1.24~T_c$ and \textit{for the first time} in full QCD at the physical point from numerical Monte Carlo simulations on the lattice above the chiral crossover. More details can be found in the main papers~\cite{Bonanno:2023ljc,Bonanno:2023thi}.
\section{Our strategy}
The sphaleron rate can be computed using the Kubo formulas as the zero frequency limit of the topological spectral function $\rho(\omega)$
\begin{equation}
\label{eq:kubo}
\Gamma_{sphal} = 2T \lim_{\omega \to 0} \frac{\rho(\omega)}{\omega},
\end{equation}
where $\rho(\omega)$ is related to the Euclidean topological charge density correlator $G(t) = \int d^3 x \braket{q(x)q(0)}$ as 
\begin{equation}
\label{eq:rho_def}
G(t) = - \int_0^{\infty} \frac{d\omega}{\pi} \rho(\omega) b(\omega,\tau),
\end{equation}
where we called $b(\omega,\tau)=\frac{\cosh\left[\omega(t-1/(2T))\right]}{\sinh\left[\omega/(2T)\right]}$.
Thus, if we are able to invert this relation, once the correlation function is known, we can extract information about the sphaleron rate. To solve this inversion problem, we apply the recently-proposed modification~\cite{Hansen:2019idp} of the Backus--Gilbert inversion method~\cite{BackusGilbert1968:aaa} called HLT method. Using this technique one looks for approximate solutions of the inverse problem~\eqref{eq:rho_def} as a smeared version of the true spectral function
\begin{equation}
\bar{\rho}(\bar{\omega}) = \int_0^\infty d\omega \Delta(\omega, \bar{\omega})\rho(\omega),
\end{equation}
where $\Delta(\omega, \bar{\omega})$ is a pseudo-gaussian smearing function defined as a linear combination of the basis function in Eq.~\eqref{eq:rho_def}, i.e.
$\Delta(\omega, \bar{\omega}) = \sum_{i=0}^{1/T} g_\tau(\bar{\omega})b(\omega,\tau)$,
being $N_\tau$ the total number of temporal data. This makes possible to find an estimate of the spectral function as 
\begin{equation}
\bar{\rho}(\bar{\omega}) = - \pi \bar{\omega} \sum_{t=0}^{1/T} g_t(\bar{\omega}) G(t).
\end{equation}
The coefficients $g_t$, that define the shape of the function $\Delta$, are fixed by minimising a functional defined as
\begin{equation}\label{eq: Functional}
F[g_t] = (1-\lambda) A_\alpha[g_t] +  \frac{\lambda}{\mathcal{C}^2}B[g_t], \,\quad  \lambda\in [0,1),
\end{equation}
where $\mathcal{C}$ is a normalization factor proportional to the correlator in a fixed point (we used $\mathcal{C}=G(tT=0.5)$ in this work) and $\lambda$ is a free parameter which is varied to check for systematics.
The functional $A_\alpha[g_t] = \int_0^{\infty} d\omega \, [\Delta(\omega,\bar{\omega}) - \delta(\omega,\bar{\omega})]^2 \,e^{\alpha \omega}$ with $\alpha < 2$, quantifies the deviation between the smearing function and a selected target function. The functional $B[g_t] = \sum_{t,t'=0}^{1/T} Cov_{t,t'} \, g_t g_{t'}$ quantifies the magnitude of the statistical uncertainties related to the final result and it is used to regularise the problem. As in~\cite{Almirante:2023wmt}, the target function is chosen as 
$\Delta(\omega, \bar{\omega}=0) = \left(\frac{2}{\sigma \pi}\right)^2 \frac{\omega}{\sinh(\omega/\sigma)}.$
The parameter $\sigma$ is crucial for determining the accuracy of the $g_t$ coefficients. In our analysis, we took several values around the value $\sigma/T\sim 1.75$, kept fixed in physical units for all the ensembles at all the temperatures, and then we performed the limit $\sigma \rightarrow 0$ using general theoretical arguments known in the literature~\cite{Bulava:2021fre,Frezzotti:2023nun,Evangelista:2023vtl}. In all cases, since the dependence is expected to start from $O(\sigma^2)$, we observed a mild dependence already around $\sigma/T=1.75$. This is an indication that this value is already sufficiently small for our precision. 
Finally, the value of the $\lambda$ parameter has been chosen inside the plateau close to $\lambda=0$. Then, another value has been selected inside the same plateau and the resulting observed systematic added to the final uncertainty.

To measure the topological charge correlation functions $G(t)$ that have been used as input for the inversion procedure, we discretised the charge density using the standard gluonic clover definition
$q_L(n) = \frac{-1}{2^9 \pi^2}\sum_{\mu\nu\rho\sigma=\pm1}^{\pm4}\varepsilon_{\mu\nu\rho\sigma}
Tr\left\{\Pi_{\mu\nu}(n)\Pi_{\rho\sigma}(n)\right\}$, where $\Pi_{\mu\nu}(n)$ is the plaquette and $\varepsilon_{(-\mu)\nu\rho\sigma} = - \varepsilon_{\mu\nu\rho\sigma}$.
By computing the time profile $Q_L(n_t)$ of the topological charge $Q_L$, we obtain the topological charge density correlator in dimensionless physical units as $\frac{G_L(tT)}{T^5} = \frac{N_t^5}{N_s^3}\braket{Q_L(n_{t,1})Q_L(n_{t,2})}$, where $N_s$ and $N_t$ are the spatial and temporal extents of the lattice and $tT=\min\left\{\vert n_{t,1} - n_{t,2} \vert/ N_t;~1-\vert n_{t,1} - n_{t,2} \vert/ N_t\right\}$ is the physical time separation between the sources entering the correlator. The topological charge profiles are calculated on smoothed configurations in order to suppress ultraviolet (UV) fluctuations. The smoothing radius is given by $r_s/a\simeq \sqrt{8n_{cool}/3}$, thus $n_{cool}/N_t^2\propto (r_sT)^2$.

\section{The quenched case}

We discretize the Euclidean pure-$SU(3)$ gauge action using the standard Wilson lattice gauge action
 $S_{\mathrm{W}} = - \frac{\beta}{3} \sum_{n,\mu>\nu}\Re Tr \left[ \Pi_{\mu\nu}(n) \right]$,
where $\beta = 6/g^2$ is the bare inverse gauge coupling and $\Pi_{\mu\nu}(n) \equiv U_\mu(n) U_\nu(n+\hat{\mu})U^\dagger_\mu(n+\hat{\nu})U^\dagger_\nu(n)$ is the plaquette.

\begin{figure}[!htb]

\begin{minipage}{.4\textwidth}
  \centering
  \includegraphics[scale=0.3]{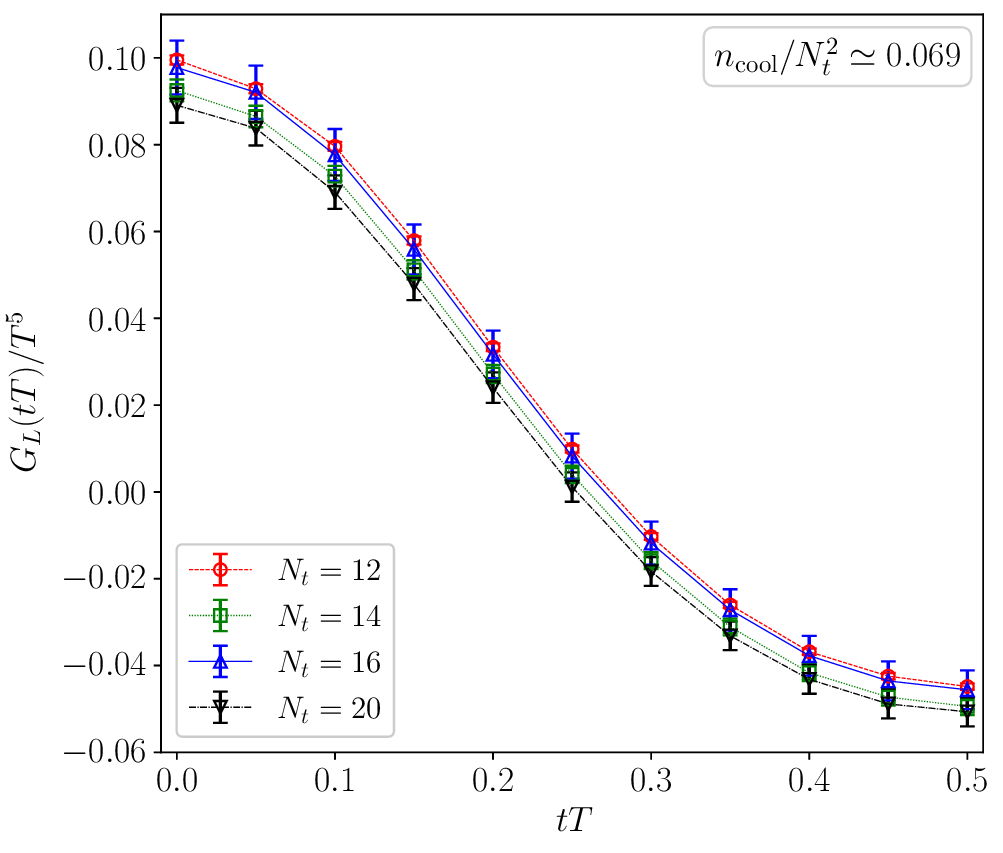}
\caption{\textit{The figure shows the correlation function $G_L(tT)$ for a fixed value of the smoothing $n_{cool}/N_t^2\simeq 0.069$ for all explored values of the lattice spacing. }}
\label{fig:tcorr_ex}
\end{minipage} \ \ \ \ \ \ \ \
\begin{minipage}{.5\textwidth}
  \centering
  \includegraphics[scale=0.3]{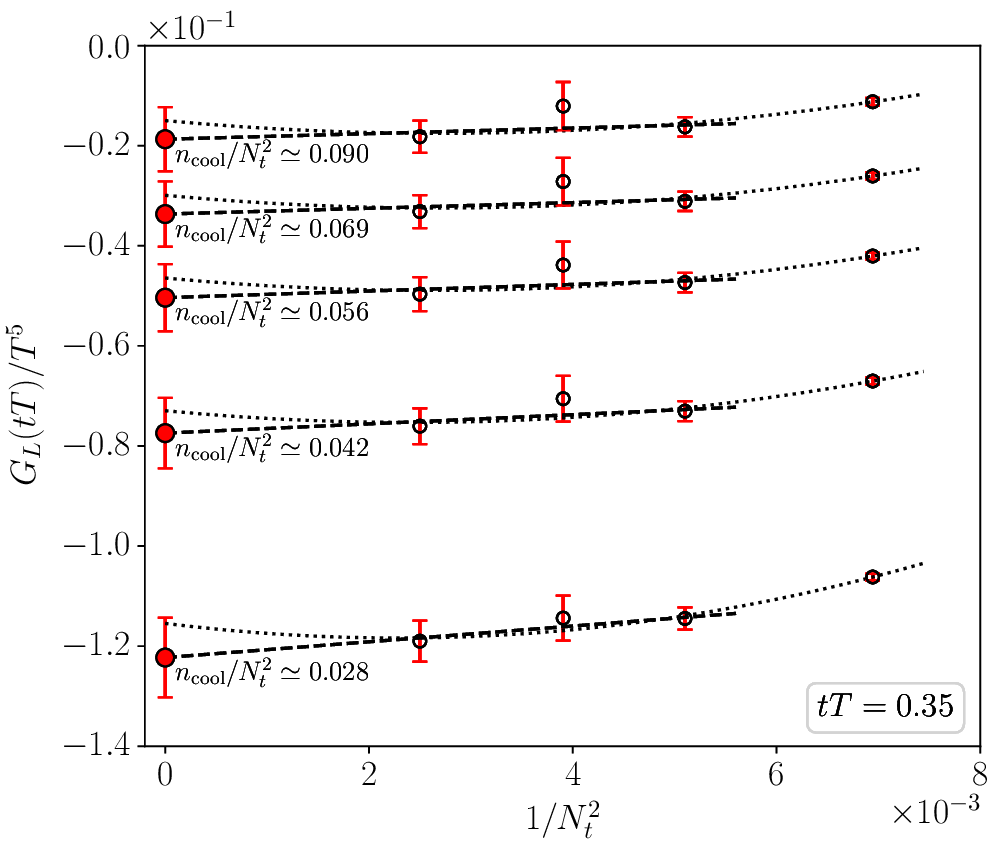}
\caption{Examples of the continuum extrapolation at fixed $n_{cool}/N_t^2$ of the correlator for two different values of $tT$.}
\label{fig:tcorr_cont_extr}
\end{minipage}
\end{figure}

All simulations parameters are summarized in Tab.~\ref{tab:simulation_summary}. We performed simulations for 4 values of $\beta$, corresponding to 4 values of the lattice spacing $a$, following a Line of Constant Physics (LCP) where the spatial volume $\left[a(\beta) N_s\right]^3 \simeq [1.66(2)~\text{fm}]^3$, the aspect ratio $N_s/N_t=3$ and the temperature $T = \left[a(\beta) N_t\right]^{-1} \simeq 357(5)~\text{MeV} \simeq 1.24(2)~T_c$ were kept fixed for each gauge ensemble. 
\begin{table}[!t]
\begin{center}
\resizebox{8cm}{!}{
\begin{tabular}{|c|c|c|c|c|c|c|}
\hline
$N_s$ & $N_t$ & $\beta$ & $a/r_0$ & $L/r_0$ & $r_0T$ & Stat.\\
\hline
36 & 12 & 6.440 & 0.09742(97) & 0.8554(86) & 3.507(35) & 80k \\
42 & 14 & 6.559 & 0.08364(84) & 0.8540(85) & 3.513(35) & 10k \\
48 & 16 & 6.665 & 0.07309(73) & 0.8551(86) & 3.508(35) & 16k \\
60 & 20 & 6.836 & 0.05846(58) & 0.8553(86) & 3.508(35) & 5k  \\
\hline
\end{tabular}
}
\end{center}
\begin{center}
\resizebox{8cm}{!}{
\begin{tabular}{|c|c|c|c|c|c|}
\hline
$N_t$ & $\beta$ & $a$~[fm] & $L$~[fm] & $T$~[MeV] & $T/T_c$ \\
\hline
12 & 6.440 & 0.04598(67) & 1.655(24) & 357.6(5.2) & 1.244(18) \\
14 & 6.559 & 0.03948(58) & 1.658(24) & 357.0(5.2) & 1.242(18) \\
16 & 6.665 & 0.03450(50) & 1.656(24) & 357.5(5.2) & 1.244(18) \\
20 & 6.836 & 0.02759(40) & 1.656(24) & 357.6(5.2) & 1.244(18) \\
\hline
\end{tabular}
}
\end{center}
\caption{Summary of simulation parameters. }
\label{tab:simulation_summary}
\end{table}
Examples of the obtained correlation functions used as inputs for our inversion algorithm are shown in Fig.~\ref{fig:tcorr_ex}.
To perform this computation we compare two strategies for extracting the rate: the first one, proposed here, involves extracting the sphaleron rate from finite lattice spacing correlators, taking the continuum limit with a fixed smoothing radius, and then performing a zero-smoothing extrapolation. The second strategy, a traditional approach, extracts the rate directly from the correlator after a double extrapolation. In both cases, the rate is obtained using a modified Backus–Gilbert procedure. The results from both strategies are compatible within errors and are compared to previous literature at similar temperatures. The new strategy yields improved results in terms of statistical and systematic uncertainties. 
\subsection{Results}
\begin{figure}[!t]
\begin{minipage}{.5\textwidth}
  \centering
  \includegraphics[scale=0.3]{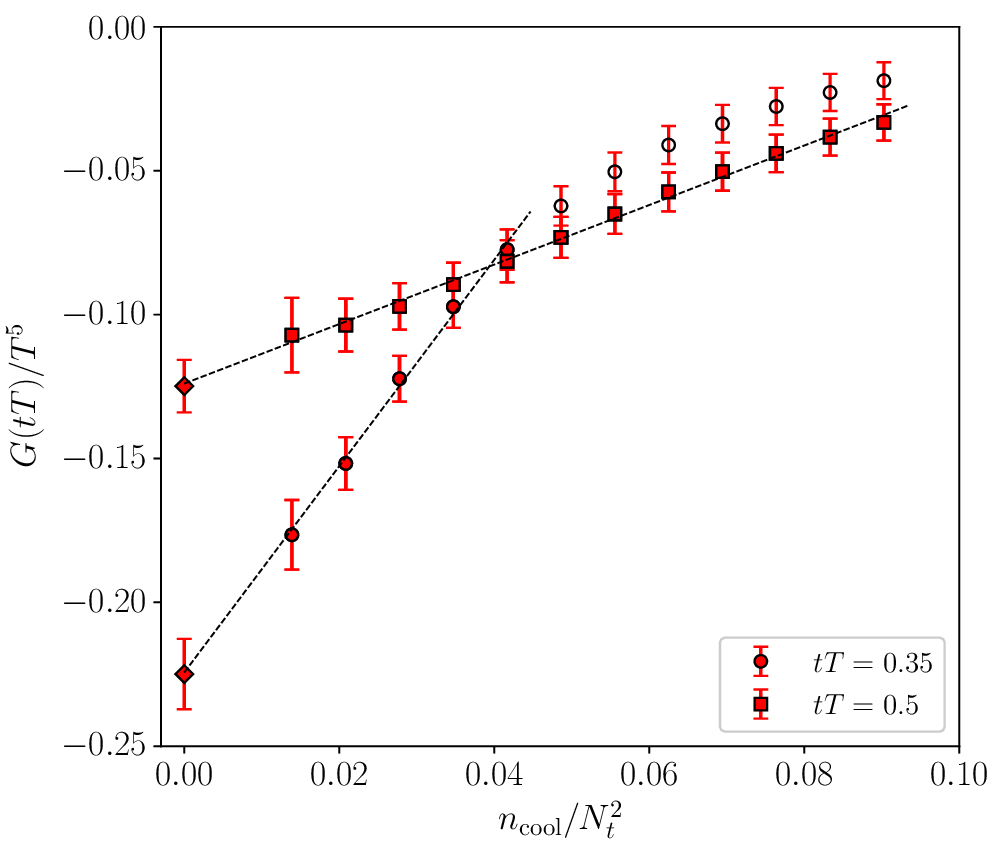}
   \captionof{figure}{Examples of the zero-cooling extrapolation of the correlator $G(tT,n_{cool}/N_t^2)$ for two different values of $tT$.}
\label{fig:tcorr_zerocool_extr}
\end{minipage} \ \ \ \ 
\begin{minipage}{.5\textwidth}
  \centering
  \includegraphics[scale=0.2]{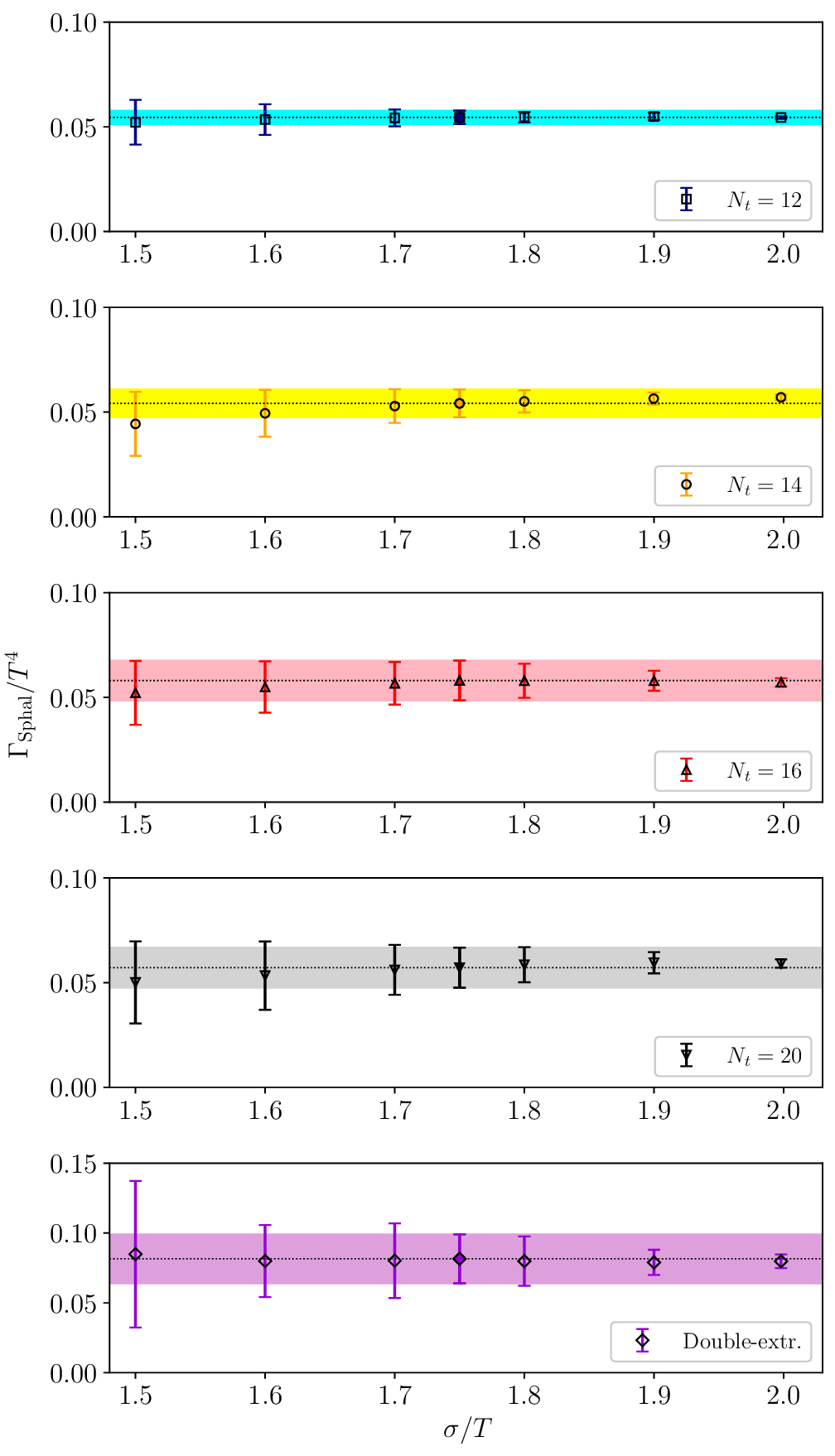}
   \captionof{figure}{Dependences as a function of $\sigma$ of the sphaleron rate. The last panel refers to the double extrapolated correlator case, while the other ones at finite lattice spacing and smoothing radius, corresponding to $n_{cool}/N_t^2 \simeq 0.04$.}
\label{fig:sigma}
\end{minipage}
\end{figure}

In this paragraph, the results obtained adopting the two methods are presented and discussed.
\subparagraph{Double-extrapolated correlators}
Using this strategy, first of all we extrapolate the correlation function $G_L(tT)/T^5$ to the continuum limit keeping fixed the smoothing radius. This is done by fixing $n_{cool}/N_t^2$ for all the lattice spacings by performing a spline cubic interpolation of our correlators at non-integer values of $n_{cool}$. The same has been done on the coarsest correlation functions in order to maintain the same physical time separation $tT$ for each lattice spacing.
To perform the continuum limit, we assume $\mathcal{O}(a^2)=\mathcal{O}(1/N_t^2)$ corrections.
After the continuum limit extrapolation, shown in Fig~\ref{fig:tcorr_cont_extr}, we can perform the zero-cooling limit $n_{cool}/N_t^2$ which is carried on assuming linear corrections in $n_{cool}/N_t^2$, see Fig.~\ref{fig:tcorr_zerocool_extr}. The result is in overall good agreement with the double-extrapolated correlator obtained for the same temperature in Ref.~\cite{Kotov:2018aaa} and is shown in Fig.~\ref{fig:tcorr_ex}.

Finally, having extracted the l.h.s. of Eq.~\eqref{eq:rho_def}, we can perform the inversion using the HLT algorithm to extract the rate $\Gamma_{sphal}$ using the strategy described in the previous Section. The bottom panel in Figure~\ref{fig:sigma} shows the $\sigma$ dependence of the sphaleron rate extracted from the double extrapolated correlation function. The dependence shows that $\sigma/T=1.75$ is already a reasonable choice, since the signal is flat in the errors as we decrease its value. 
Our final estimate is
\begin{equation}
\frac{\Gamma_{sphal}}{T^4} = 0.079(25), \ \ \ T = 1.24T_c.
\end{equation}
This result is compatible with the one reported in Ref.~\cite{Kotov:2018aaa}, although the central value is $\sim 33\%$ smaller.

\begin{figure}
\begin{minipage}{.5\textwidth}
  \centering
  \includegraphics[scale=0.3]{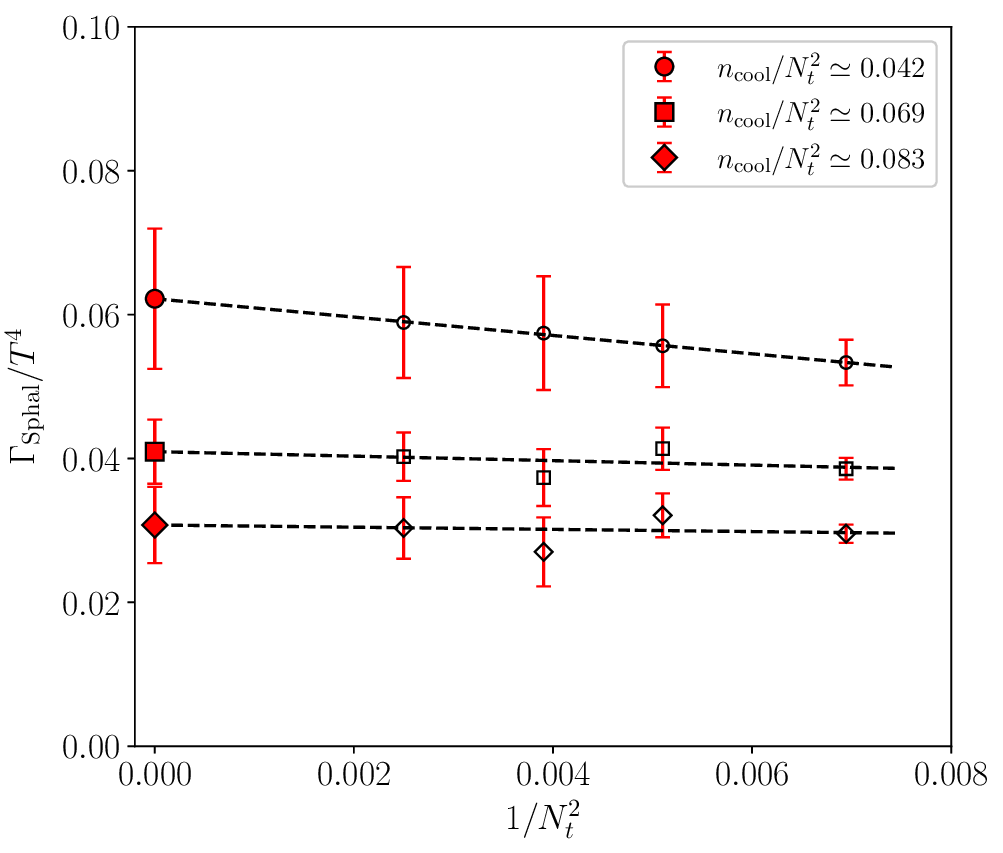}
  \captionof{figure}{Continuum extrapolation of the sphaleron rate.}
  \label{fig:Continuum_Spha}
\end{minipage}\ \ \ \ \ \ 
\begin{minipage}{.5\textwidth}
  \centering
  \includegraphics[scale=0.3]{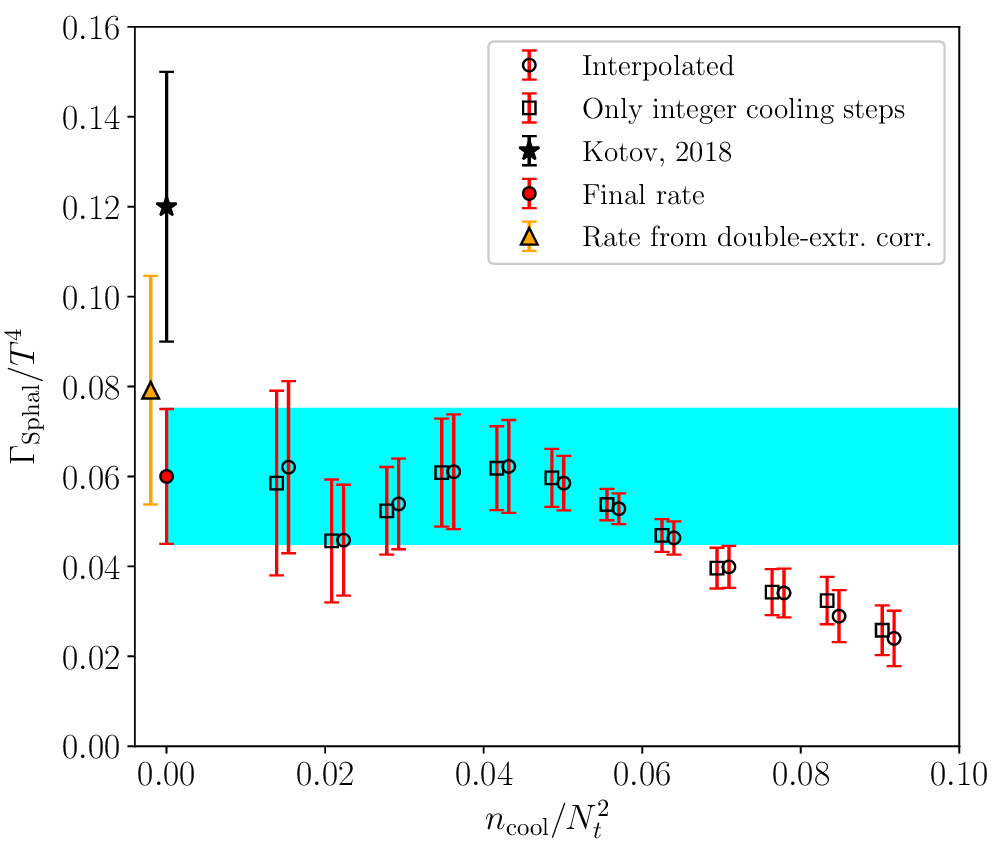}
   \captionof{figure}{Zero cooling radius extrapolation.}
\label{fig:Zero_Cool_Spha}
\end{minipage}
\end{figure}

\paragraph{Double extrapolated sphaleron rate.}
Using the second approach, we perform the inversion directly on the correlation functions at finite lattice spacing and finite smoothing radius. In Fig.~\ref{fig:sigma} it is shown the dependence in $\sigma$. The results for the sphaleron rate are firstly extrapolated in the continuum at fixed $n_{cool}/N_t^2$ assuming $\mathcal{O}(a^2)$ corrections, see Fig.~\ref{fig:Continuum_Spha}. Then, the zero smoothing radius limit is performed on the continuum extrapolations of the sphaleron rate, as shown in Fig.~\ref{fig:Zero_Cool_Spha}.
Our final estimate using this method is 
\begin{equation}
\label{eq:final_result_rate}
\frac{\Gamma_{sphal}}{T^4} = 0.060(15), \qquad T \simeq 1.24~T_c,
\end{equation}
which is in perfect agreement with the one obtained using the previous method, but is more accurate. Moreover, also this result presents a smaller central value for the sphaleron rate compared to the one reported in Ref.~\cite{Kotov:2018aaa} at the same temperature, $\Gamma_{sphal}/T^4=0.12(3)$, even if it is still compatible with it.
Furthermore, if we compare our result with Ref.~\cite{BarrosoMancha:2022mbj}, where a completely different strategy was used, the smallest temperature result in that work, for $T \simeq 1.3~T_c$, close to our measure, is in perfect agreement with the one extracted in this work.
\section{The full QCD case}
\begin{table}[!htb]
\begin{center}
\resizebox{7cm}{!}{
\begin{tabular}{ |c|c|c|c|c|c|c|}
\hline
$T$~[MeV] & $T/T_c$ & $\beta$ & $a$~[fm] & $a m_s \cdot 10^{-2}$ & $N_s$ & $N_t$ \\
\hline
\multirow{5}{*}{230} & \multirow{5}{*}{1.48} & 3.814* & 0.1073 & 4.27 & 32 & 8  \\
&& 3.918* & 0.0857 & 3.43 & 40 & 10 \\
&& 4.014  & 0.0715 & 2.83 & 48 & 12 \\
&& 4.100  & 0.0613 & 2.40 & 56 & 14 \\
&& 4.181  & 0.0536 & 2.10 & 64 & 16 \\
\hline	
\multirow{4}{*}{300} & \multirow{4}{*}{1.94} & 3.938 & 0.0824 & 3.30 & 32 & 8  \\
&& 4.059  & 0.0659 & 2.60 & 40 & 10 \\
&& 4.165  & 0.0549 & 2.15 & 48 & 12 \\
&& 4.263  & 0.0470 & 1.86 & 56 & 14 \\
\hline	
\multirow{4}{*}{365} & \multirow{4}{*}{2.35} & 4.045 & 0.0676 & 2.66 & 32 & 8 \\
&& 4.175  & 0.0541 & 2.12 & 40 & 10 \\
&& 4.288  & 0.0451 & 1.78 & 48 & 12 \\
&& 4.377  & 0.0386 & 1.55 & 56 & 14 \\
\hline
\multirow{4}{*}{430} & \multirow{4}{*}{2.77} & 4.280 & 0.0458 & 1.81 & 32 & 10 \\
&& 4.385 & 0.0381 & 1.53 & 36 & 12 \\
&& 4.496 & 0.0327 & 1.29 & 48 & 14 \\
&& 4.592 & 0.0286 & 1.09 & 48 & 16 \\
\hline
\multirow{3}{*}{570} & \multirow{3}{*}{3.68} & 4.316  & 0.0429 & 1.71 & 32 & 8 \\
&& 4.459  & 0.0343 & 1.37 & 40 & 10 \\
&& 4.592  & 0.0286 & 1.09 & 48 & 12 \\
\hline
\end{tabular}
}
\end{center}
\caption{Summary of simulation parameters. See Ref.~\cite{Athenodorou:2022aay} for more details on the configuration generation.}
\label{tab:simul_params}
\end{table}
Finally, we extended our computation to full QCD. We did Monte Carlo simulations of $N_f=2+1$ QCD at the physical point across five temperatures: $T=230, 300, 365, 430,$ and $570$ MeV. For each temperature, we explored 3-5 lattice spacing values while maintaining a constant physical lattice volume. We selected the bare coupling and quark masses to stay on a Line of Constant Physics (LCP), where $m_s/m_l = 28.15$ and $m_\pi\simeq135$ MeV were kept fixed at their physical values~\cite{Aoki:2009sc,Borsanyi:2010cj,Borsanyi:2013bia}. The gauge sector was discretised using the tree-level Symanzik improved Wilson gauge action, and the quark sector employed rooted stout staggered fermions. In Tab~\ref{tab:simul_params} the simulation parameters are summarised.

\begin{figure}
\begin{minipage}{.5\textwidth}
  \centering
  \includegraphics[scale=0.3]{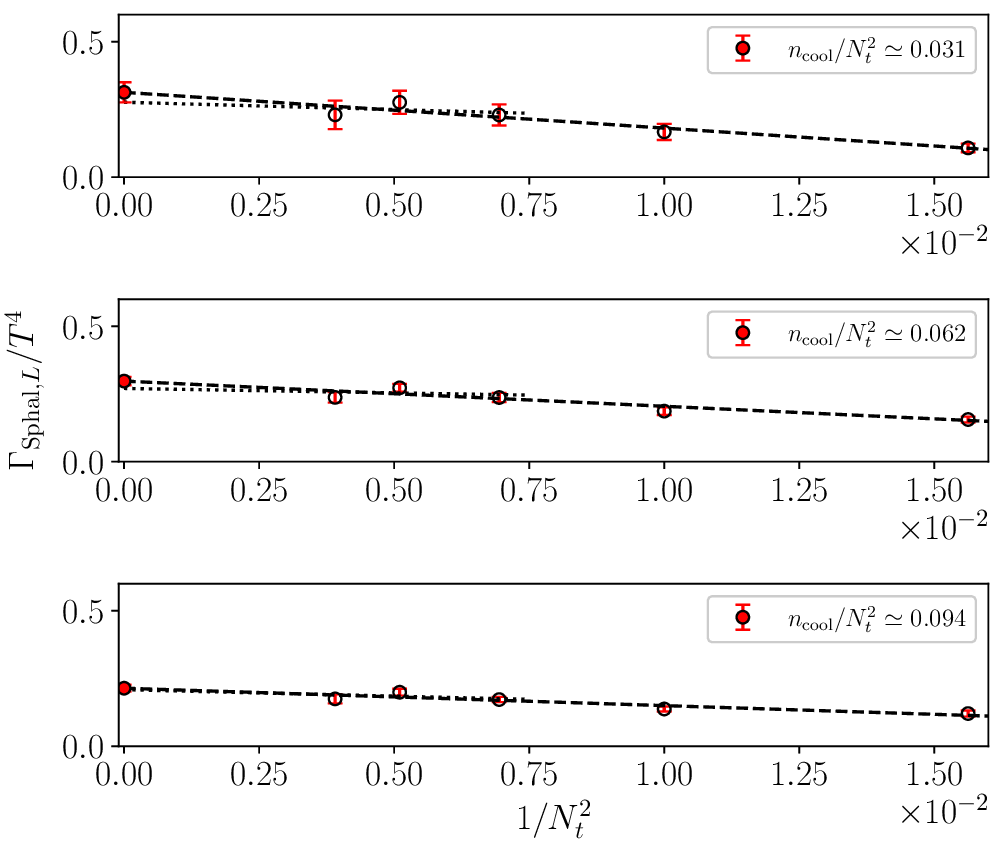}
  \captionof{figure}{Continuum extrapolation of the sphaleron rate.}
  \label{fig:Continuum_FullQCD}
\end{minipage}\ \ \ \ \ \ 
\begin{minipage}{.5\textwidth}
  \centering
  \includegraphics[scale=0.3]{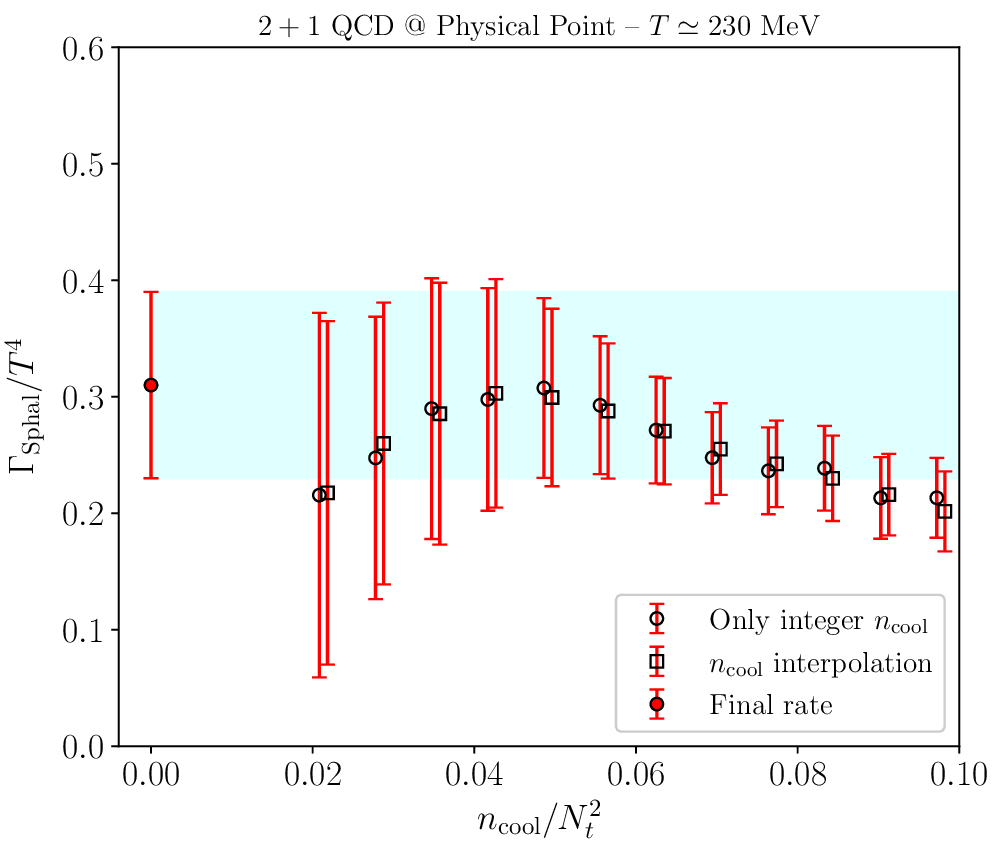}
   \captionof{figure}{Zero cooling radius extrapolation.}
\label{fig:Zero_Cool_Spha_FullQCD}
\end{minipage}
\end{figure}

Here, we followed the same procedure as before performing the double limit directly on the sphaleron rate. In Figs.~\ref{fig:Continuum_FullQCD}-\ref{fig:Zero_Cool_Spha_FullQCD}, we show the extrapolation to the continuum limit and at zero cooling radius for the temperature $T=230$ MeV. In Table~\ref{tab:rate_res}, we summarize our results for the sphaleron rate as a function of the temperature.

\begin{figure}[!t]

\begin{minipage}{.5\textwidth}
  \centering
  \includegraphics[scale=0.3]{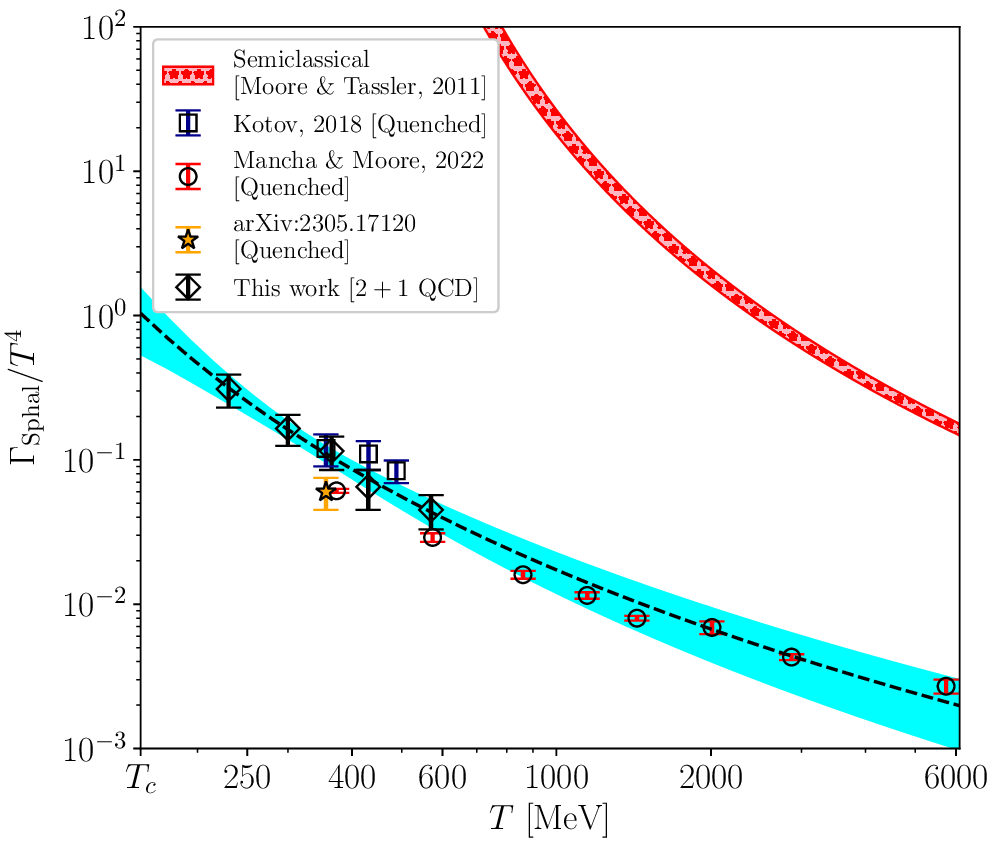}
\end{minipage} 
\begin{minipage}{.5\textwidth}
  \centering
  \includegraphics[scale=0.3]{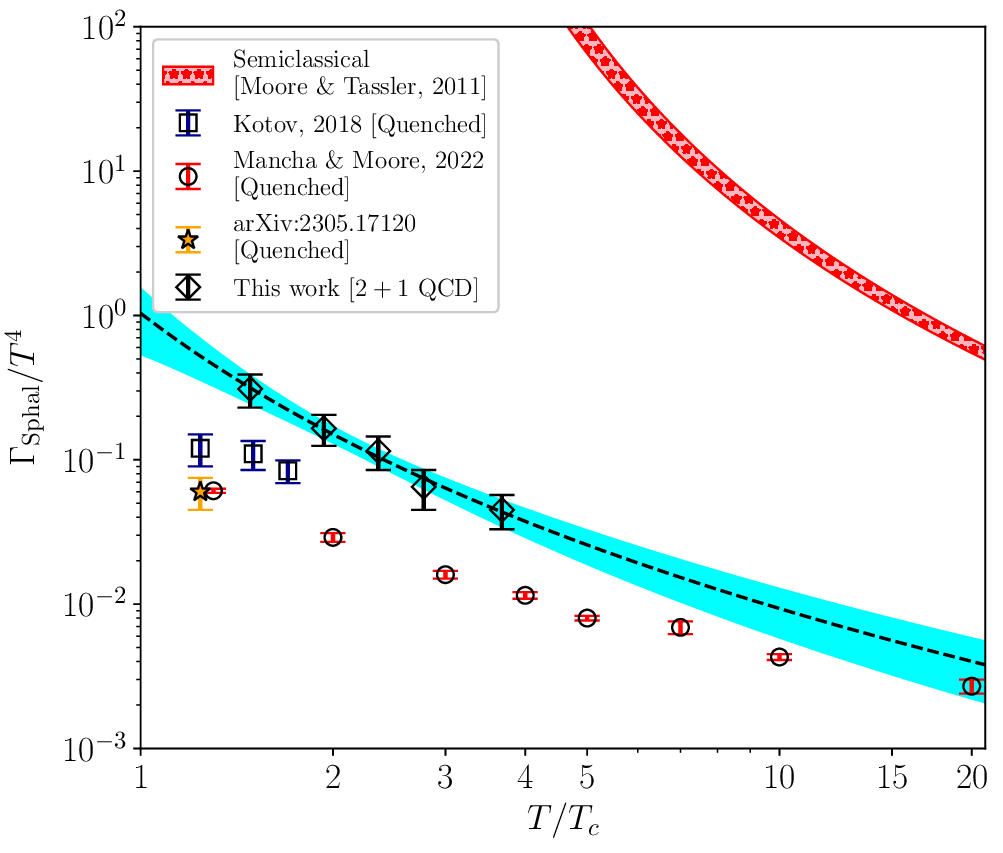}
\end{minipage}
\caption{Sphaleron rate for 2+1 full QCD at the physical point as a function of temperature T (diamond points). Dashed line and uniform shaded area represent best fit of our results. Starred shaded area depicts semiclassical prediction.}
\label{fig:fit1}

\end{figure}

Let us comment on those results. In Fig.~\ref{fig:fit1}, we show our comparison with our pure gauge results and the previous determinations~\cite{Kotov:2018aaa,BarrosoMancha:2022mbj}. 
These are valid in two different limits of energy scales. The full QCD determinations turn out to be slightly larger (although of the same order of magnitude) than the quenched ones, both when we report the rates in terms of the absolute $T$ in MeV, and when we report them in terms of $T/T_c$. Furthermore, we tried to look for ans\"atze that could describe the behaviour of our results.
\begin{table}[!htb]
\begin{center}
\begin{tabular}{ |c|c|}
\hline
&\\[-1em]
$T$~[MeV] & $\Gamma_{sphal}/T^4$ \\
&\\[-1em]
\hline
230 & 0.310(80)\\
300 & 0.165(40)\\
365 & 0.115(30)\\
430 & 0.065(20)\\
570 & 0.045(12)\\
\hline
\end{tabular}
\end{center}
\caption{Summary of the determinations of the sphaleron rate of $2+1$ full QCD at the physical point.}
\label{tab:rate_res}
\end{table}

From Refs.~\cite{Moore:2010jd,Berghaus:2020ekh}, we refer to the semiclassical estimate for the sphaleron rate as $\frac{\Gamma_{sphal}}{T^4} \simeq C_1 \alpha_s^5$, where $\alpha_s$ is the running strong coupling. Thus, we used as ansatz
\begin{equation}\label{eq:fit_sphal_ans1}
\frac{\Gamma_{sphal}}{T^4} = \left[\frac{A}{\log(T^2/T^2_c) + \log(B^2)}\right]^5,
\end{equation}
where we used the 1-loop result for the temperature running of $\alpha_s(T)$. 
The fit is shown in Fig~$\ref{fig:fit1}$ as a dashed line while the uniform shaded area represents the corresponding error band; the fit parameters turn out to be $A=2.96(51)$ and $B=4.3(1.7)$. The red band shows the semiclassical prediction results using the coefficients predicted by the theory, namely $B_0 = T_c/\Lambda_{QCD} \simeq 0.46(2)$ using the latest world-average FLAG value for the 3-flavor dynamically-generated scale $\Lambda_{QCD}^{(\overline{\mathrm{MS}})}(\mu = 2~\mathrm{GeV}) \simeq 338(12)$~MeV~\cite{FlavourLatticeAveragingGroupFLAG:2021npn}, and $A_0 =C_1^{1/5}C_2 \simeq 3.08(2)$ using the expressions for $C_1$ and $C_2$ reported, respectively, in Refs.~\cite{Moore:2010jd,Steffens:2004sg}. While $A$ is in good agreement with the prediction $A_0$, the pole parameter $B$ is larger by an order of magnitude compared to $B_0$. As a final remark, we would also like to mention that, despite a semiclassically-inspired logarithmic power-law fits well our full QCD results for the sphaleron rate, also other functional forms could describe the $T$-behavior of our data, like a simple power-law in T. More details can be found in the main paper~\cite{Bonanno:2023thi}. 
\section{Conclusions}
We computed the sphaleron rate from Euclidean lattice correlators of the topological charge by solving an inverse problem. The method that we used is the recently-introduced HLT method~\cite{Hansen:2019idp}. Our strategy for the computation of the sphaleron rate has been firstly tested on the pure gauge case~\cite{Bonanno:2023ljc} and then extended for the first full QCD computation~\cite{Bonanno:2023thi}. The main strategy, used in both cases, has been to invert the finite-lattice spacing and finite-smoothing radius correlators, postponing the double-extrapolation directly on the final quantity, i.e. the sphaleron rate. In pure gauge our result is in agreement with the ones already in the literature, while in full QCD, having studied the temperature behaviour, we tried to describe our data using semiclassically-inspired functional form. However, also other functional forms, such as a regular power-law decay of the rate, are shown to describe well our data~\cite{Bonanno:2023thi}. 

In the future, it would be extremely interesting to repeat our calculation of the sphaleron rate adopting a different fermionic discretisation and also investigate higher temperatures in order to better clarify the actual temperature behaviour. Finally, it would be interesting to extend present computations to the case of non-zero spatial momentum $\vec{k}$.

\newpage
\bibliographystyle{JHEP}
\bibliography{biblio}

\end{document}